\documentclass[sigconf]{acmart}
\usepackage{graphicx}
\usepackage{amsmath}
\usepackage{booktabs}
\usepackage{algorithm}
\usepackage{algorithmic}
\usepackage{amsfonts}
\usepackage{multirow}
\usepackage{makecell}
\usepackage{subcaption}
\usepackage{color}
\usepackage{bm}
\usepackage{epstopdf}
\usepackage{url}
\usepackage[cal=cm]{mathalfa}
\usepackage{balance}
\usepackage{threeparttable}
\usepackage{lipsum}
\usepackage{enumitem}
\usepackage{pifont}
\usepackage{tabularx}
\usepackage{makecell}
\usepackage{float}
\usepackage[table]{xcolor}
\usepackage{array}
\usepackage{afterpage}
\usepackage{tcolorbox}

\definecolor{mydeepgreen}{RGB}{0, 100, 0} 

\newcommand{\myred}[1]{\textcolor{red}{#1}}
\newcommand{\mygreen}[1]{\textcolor{mydeepgreen}{#1}}    
\newcommand{\myorange}[1]{\textcolor{orange}{#1}}  

\setlist[itemize]{leftmargin=*}

\AtBeginDocument{%
  \providecommand\BibTeX{{%
    \normalfont B\kern-0.5em{\scshape i\kern-0.25em b}\kern-0.8em\TeX}}}

\renewcommand\footnotetextcopyrightpermission[1]{}

\author{Tian Xia*, Jiaqi Zhang*, Yueyang Liu*, Hongjian Dou*, Tingya Yin*, Jiangxia Cao$^\star$, Xulei Liang, Tianlu Xie, Lihao Liu, Xiang Chen, Shen Wang, Changxin Lao, Haixiang Gan, Jinkai Yu, Keting Cen, Lu Hao, Xu Zhang, Qiqiang Zhong, Zhongbo Sun, Yiyu Wang, Shuang Yang, Mingxin Wen, Xiangyu Wu, Shaoguo Liu, Tingting Gao, Zhaojie Liu, Han Li, Kun Gai}
\thanks{* Equal Contributions, Jiangxia Cao is the corresponding author.}
\affiliation{
  \institution{Kuaishou Technology, Beijing, China}
 \country{\{xiatian06, zhangjiaqi15, liuyueyang05, douhongjian, yintingya, caojiangxia, liangxulei, xietianlu, liulihao, chenxiang08, wangshen, laochangxin, ganhaixiang03, yujinkai, cenketing03, luhao, zhangxu20, lifan11, liming28, liubin21, yangshuang08, wenmingxing, wuxiangyu06, liushaoguo, lisize, zhaotianxing, lihan08\}@kuaishou.com, gai.kun@qq.com}
}

\begin{document}
\title{QARM V2: Quantitative Alignment Multi-Modal Recommendation for Reasoning User Sequence Modeling}

\renewcommand{\shorttitle}{QARM V2}

\begin{abstract}
With the evolution of large language models (LLMs), there is growing interest in leveraging their rich semantic understanding to enhance industrial recommendation systems (RecSys). Traditional RecSys relies on ID-based embeddings for user sequence modeling in the General Search Unit (GSU) and Exact Search Unit (ESU) paradigm, which suffers from low information density, knowledge isolation, and weak generalization ability. While LLMs offer complementary strengths with dense semantic representations and strong generalization, directly applying LLM embeddings to RecSys faces critical challenges: representation unmatch with business objectives and representation unlearning end-to-end with downstream tasks.
In this paper, we present QARM V2, a unified framework that bridges LLM semantic understanding with RecSys business requirements for user sequence modeling. 
At the GSU side, we introduce a reasoning-based item alignment mechanism that generates business-aligned LLM embeddings, enabling semantic-aware retrieval of relevant historical subsequences. 
At the ESU side, we propose Res-KmeansFSQ, a hybrid quantization method that combines Residual K-means with Finite Scalar Quantization to produce multi-level Semantic IDs (SIDs) with minimal code conflicts. These SIDs serve as learnable discrete features that can be optimized end-to-end with the ranking model.
Extensive offline experiments and online A/B tests on Kuaishou's short-video and live-streaming platforms demonstrate significant improvements over existing methods, validating its effectiveness.
\end{abstract}

\maketitle

\section{Introduction}
Kuaishou, as the leading short-video and live-streaming distribution platform in China, attracts more than 400 million daily active users who share their lifestyles, interact with fans, and buy/sell products, providing a vibrant economic ecosystem.
Among Kuaishou, tens of millions of fresh new items are uploaded every day (e.g., short-videos, live-streams, products, etc.), making it crucial to discover the right items that match our users' interests.
To achieve this goal, a powerful industrial Recommendation System (RecSys)~\cite{kuaiformer, youtubednn, home} is a vital engine that drives user engagement and revenue growth to support our business.

\begin{figure}[t!]
  \centering
  \includegraphics[width=8cm,height=10cm]{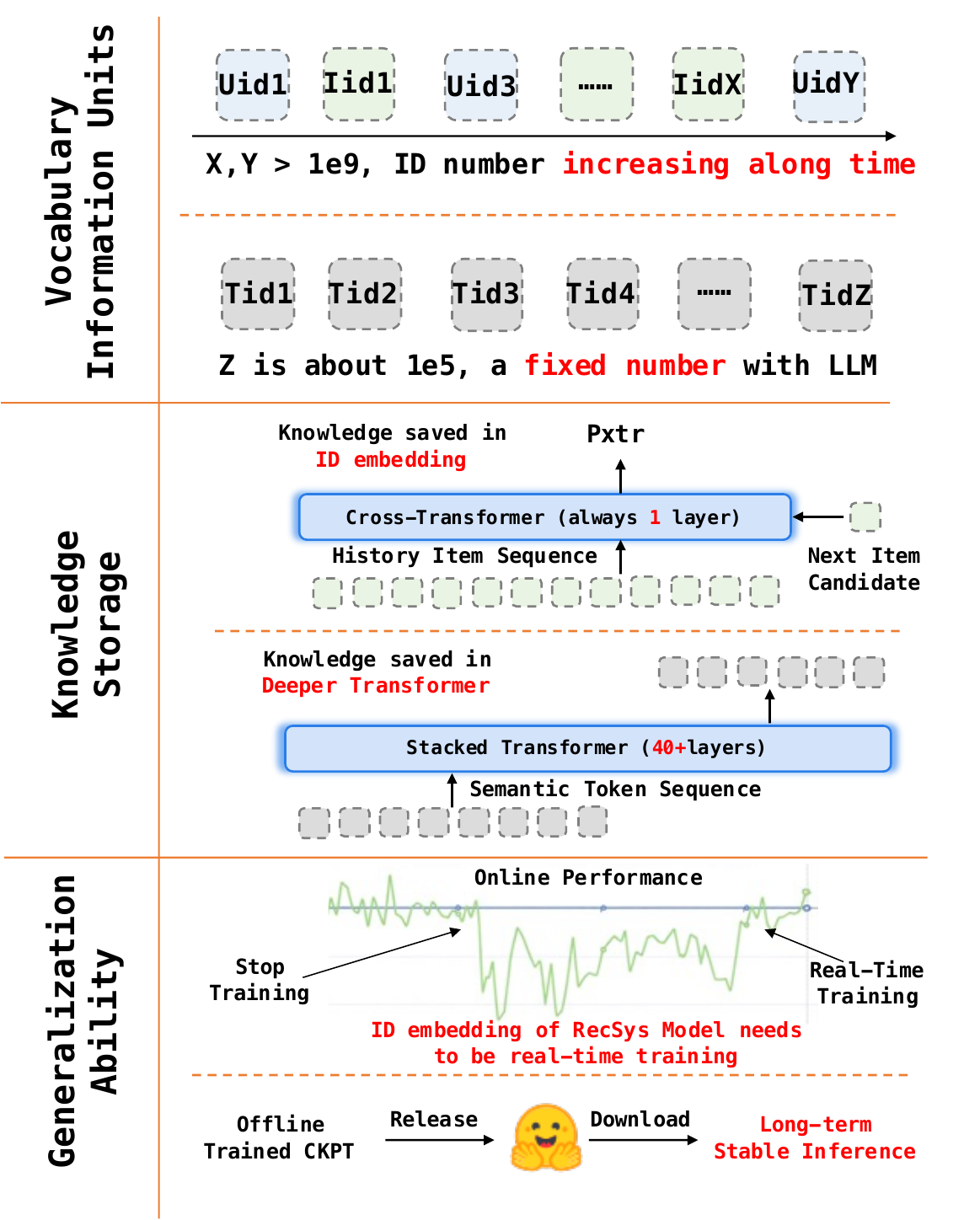}
    \caption{ID-based RecSys Model and LLM Differences.}
    \label{fig:background}
    \vspace{-0.4cm}
\end{figure}

\begin{figure}[t!]
  \centering
  \includegraphics[width=8cm,height=4.4cm]{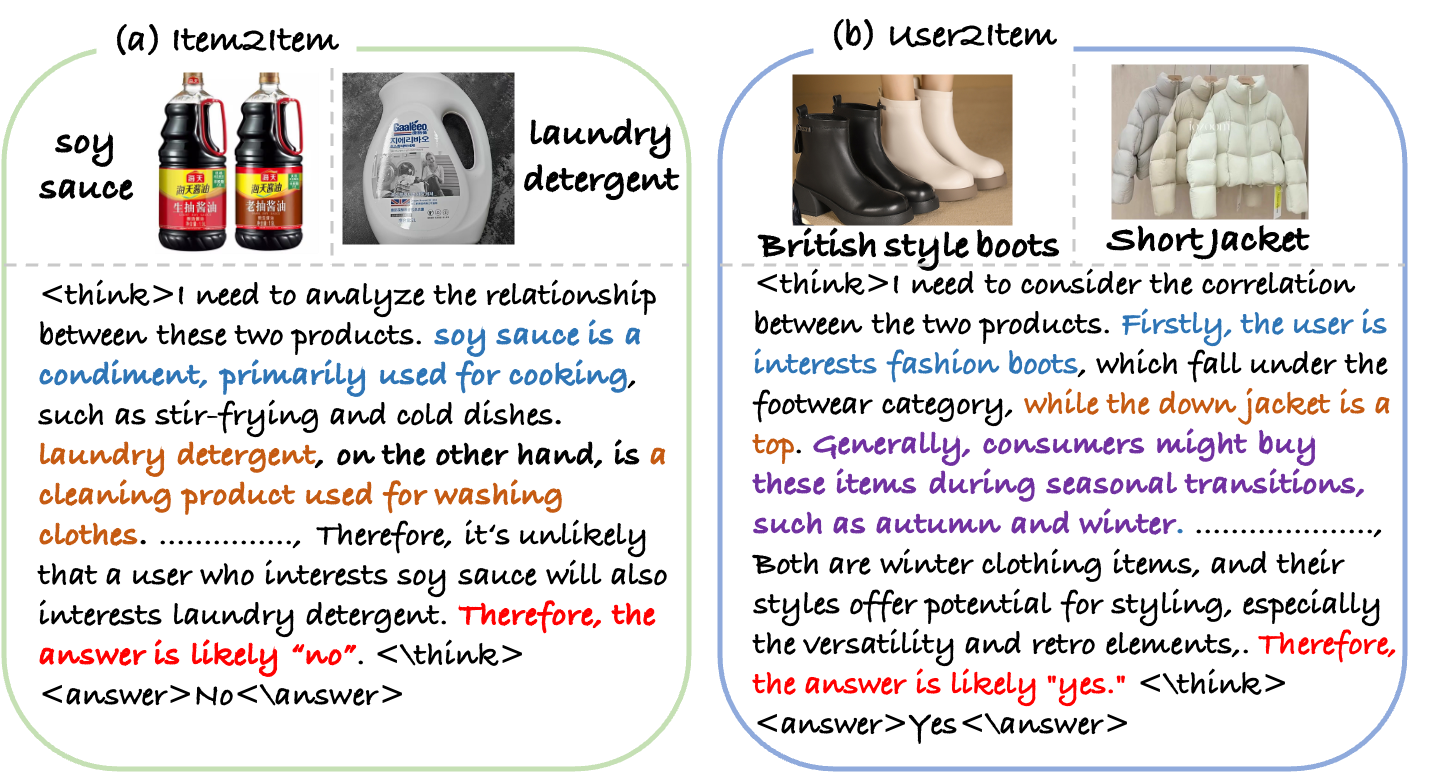}
    \caption{LLM understanding reasoning ability is necessary to filter the noise item pairs behind RecSys model: find the unrelated hot-popular biased item pairs from Item2Item model, and identify the different but relevant item pairs from User2Item model.}
    \label{fig:reasoing}
\end{figure}

In broader sense, user sequence modeling~\cite{zhou2018deep,sasrec} is one of the most important components in industrial RecSys, which is a crucial scaling direction~\cite{chai2025longer} to capture the user's dynamic preferences and behaviors, contributing convincing gains in real businesses services.
On our platform, the accumulated interaction sequence exceeds 100,000 for each active user, which strongly reflects the user's long-term interest evolution and short-term interest shifts.
However, directly inputting the entire 100,000 user sequence to a RecSys model in a streaming training setting will introduce massive offline computation pressure, making it hard to afford such training resources.
To alleviate computation, existing methods typically adopt a two-stage strategy to revisit the user interaction sequence and for each next target item candidate~\cite{pi2020search, pi2019practice}:
\begin{itemize}
    \item Given one item candidate, we first use a simple (not optimized directly) \textbf{General Search Unit (GSU)} to review users' lifelong history and then filter top-K related item subsequences.
    \item According to the subsequences, we next use a complex (optimized end-to-end) \textbf{Exact Search Unit (ESU)} to compress subsequence information and obtain fine-grained user interests.
\end{itemize}
Although the above GSU/ESU paradigm successfully extends the boundary of RecSys, but the latest efforts are rely on the user/item ID embeddings to support both GSU/ESU similarity calculation, e.g., TWIN~\cite{chang2023twin}, TWIN V2~\cite{si2024twin}.
This reliance on ID embeddings introduces inherent barriers, resulting in the following trade-offs:
\begin{itemize}
    \item \mygreen{\textbf{Low Information Units}}: user preferences and item properties are learned via user-item interaction logs, which could capture coarse semantics (e.g., item tags),
    but suffer from the long-tail problem (e.g., new item hard to beat old hot items) and limited fine-grained semantic expressiveness (e.g., what the item means).
    \item \mygreen{\textbf{Knowledge Isolation}}: the collaborative signals are saved in billion-scale item embeddings separately; once an item is no longer distributed, its knowledge will be discarded completely.
    \item \mygreen{\textbf{Weak Generalization Ability}}: industrial RecSys models are desired to chase the latest interaction data; once they stop streaming training, online performance will drop sharply.
\end{itemize}
With those inherent weaknesses in ID-based information units, it is reasonable to incorporate multi-modal language models~\cite{yang2025kwai,team2025kwai}, hoping the learned transferring world-knowledge semantics could enable more precise GSU/ESU calculations. 
Indeed, the LLM~\cite{achiam2023gpt,bai2023qwen} actually has desired characteristics, as shown in Figure~\ref{fig:background}:%
\begin{itemize}
\item \myred{\textbf{Dense Information Units}}: different from massive user/item IDs in RecSys, the LLM only utilizes fewer than 20,000 token IDs vocabulary to express world knowledge in human society.
\item \myred{\textbf{Knowledge Integrated}}: With deeper and larger Transformer parameters that store knowledge, LLMs can understand token sequence semantics accurately. 
\item \myred{\textbf{High Generalization Ability}}: Once an LLM is trained, it can provide long-term stable inference; for any token sequence, it can be accurately understood to explain for us.
\end{itemize}

As discussed above, LLMs can effectively complement the deficiencies of ID-based features by providing ideal auxiliary information for GSU/ESU~\cite{wu2025muse}. 
To our knowledge, a widely used approach in industrial companies is to first use LLMs to generate representations for each item, and then utilize these cached embeddings as additional features for both GSU and ESU.
In practice, its ceiling performance is limited if the LLM is not tuned alignment task with downstream RecSys.
\myorange{In other words, directly adding pre-trained multimodal embeddings as additional item features brings very limited gains, since our goal is not to recommend lookslike similar items our users have already seen, but rather to discover new items of interest expanded from their historical behaviors.}
For this phenomenon, the common wisdom holds the following perspective:
\begin{itemize}
  \item \textit{Representation Unmatch} (LLM generated embedding not match RecSys demand): The objectives of pre-trained LLMs (such as image captioning and question answering) are inconsistent with RecSys objectives (image and text matching vs. user click prediction). The representations provided by LLMs may not be what downstream businesses need. For example, toothpaste and ointment may appear similar due to their tubular packaging, but serve completely different usage scenarios.
  \item \textit{Representation Unlearning} (Freezing LLM embeddings is hard to adopt in dynamic RecSys): Pre-trained representations cannot be learned end-to-end, stop being updated through gradients to adapt to downstream tasks, making them prone to incompatibility and limiting their flexibility.
\end{itemize}

\begin{figure}[t!]
  \centering
  \includegraphics[width=8cm,height=4.4cm]{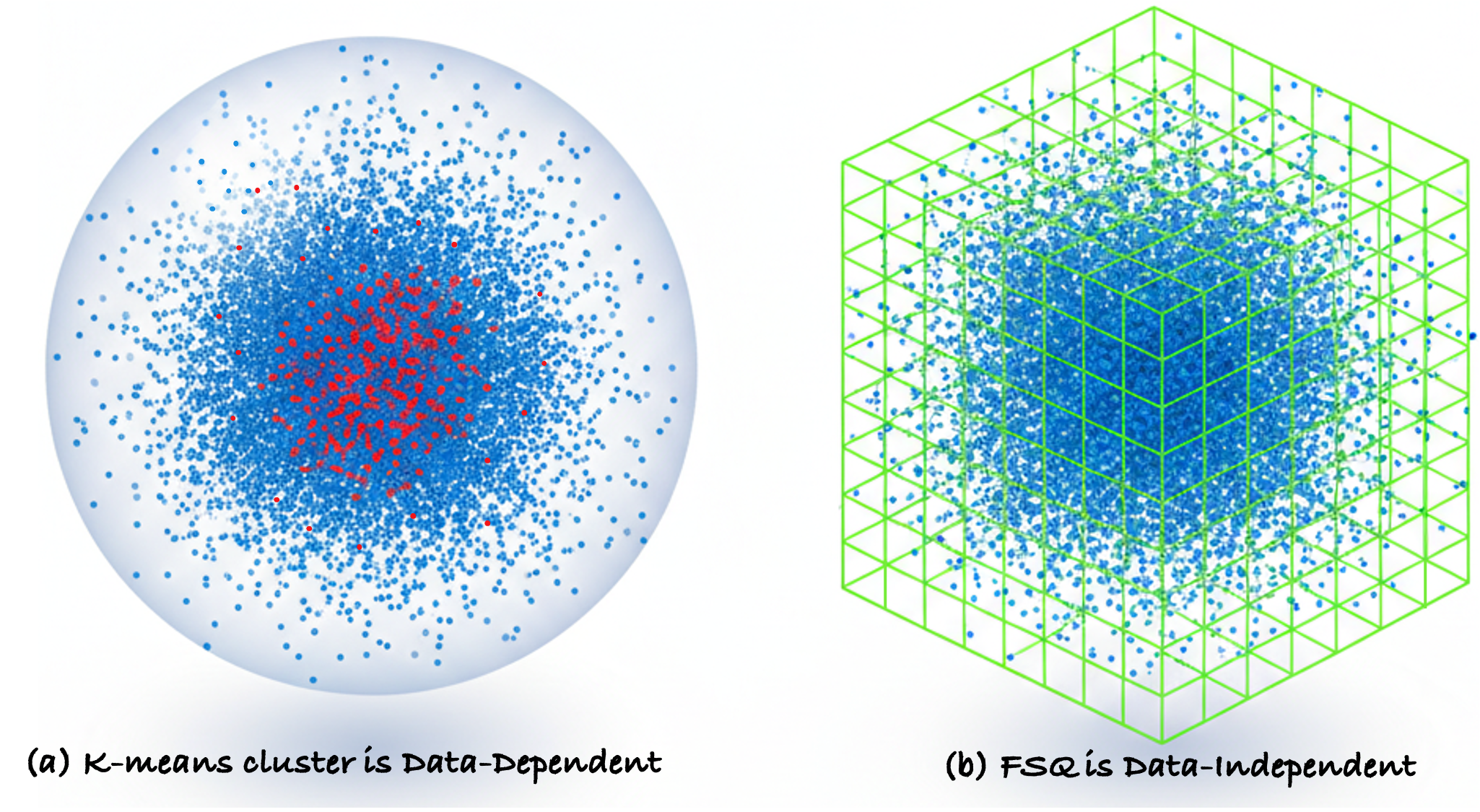}
    \caption{The K-means algorithm is a distribution-dependent quantification method that is highly affected by the training data; the distribution-edge points will be pulled closer to the distribution-dense area. FSQ is a distribution-independent quantification method that keeps each data point's position.}
    \label{fig:quant}
\end{figure}

Can we find a way that fully unleashes the LLM potential in lifelong user interest modeling? 
\myorange{That LLM-generated item embeddings could reflect appropriate business semantics on the GSU side, and enable friendly learning for better adaptation on the ESU side?}
In this paper, we present QARM V2 to provide RecSys business-aligned LLM embeddings and fine-grained Semantic IDs~\cite{rajput2023recommender, deng2025onerec}, especially designed for lifelong sequence modeling.
In summary, at the GSU side, our QARM V2 introduces a more comprehensive reasoning item alignment mechanism to enhance representation consistency in both business and world knowledge.
At the ESU side, we propose the Res-KmeansFSQ quantitative code mechanism to generate learnable Semantic IDs with minor code conflict for more accurate optimization.

\textbf{Reasoning Item Alignment Mechanism}: To alleviate the Representation-Unmatch problem, we consider fine-tuning a pre-trained LLM to understand the similarity logic in business scenarios.
To reach this goal, the previous effort QARM~\cite{luo2024qarm} proposes a naive item alignment mechanism, which exports the learned high similarity (trigger Item, target Item) pairs from RecSys model as contrastive supervision signals:
\begin{itemize}
\item \textit{Item2Item Swing retrieval model} exported item pairs describe the stable exploitable item relationships that have been accumulated in our RecSys (most users clicked product A and product B)\cite{yang2020large}.
    \item \textit{User2Item Two-Tower retrieval model} exported item pairs express unstable exploring item relationships (the current user clicked product A and product B).
\end{itemize}
We acknowledge those item pairs are vital to reflect the exploration-exploitation character in our platform business environment; however, we also find some item pairs are noisy or biased and should be further filtered.
As shown in Figure~\ref{fig:reasoing}, the Item2Item retrieval model easily provides biased hot-popular item pairs, such as the (soy sauce, laundry detergent) pair; they are highly affected by exposure bias but have no underlying logic between the products themselves.
Besides, the item pairs exported by the User2Item retrieval model are highly flexible.
Some may be totally unrelated in terms of modal content and underlying logic, but some share similar usage scenarios and underlying connections, e.g., toothpaste and dental floss.
In QARM V2, we argue that it is necessary to apply LLMs to filter the generated item pairs provided by RecSys models, rejecting noisy samples.
Based on our observations, we deploy the latest pre-trained reasoning LLMs to judge whether the item pairs are closely related in world knowledge.
Thereby we could obtain the more comprehensive item alignment data that actually happen on our platform, to balance exploration and exploitation with deeper underlying real-world logic.

Besides, to not hurt the LLM pre-training paradigm to much, we devise a three-segment trick to conduct the item pair contrastive and the question-answer generation at same time, further enhancing training stability to transform GPTs as an embedding generator.

\textbf{Res-KmeansFSQ Quantitative Code Mechanism}:
To overcome the Representation-Unlearning problem, the previous QARM proposes a Res-Kmeans algorithm to quantify LLM embeddings into multi-level residual Semantic IDs, thereby the downstream RecSys model can assign additional embeddings for end-to-end optimization~\cite{zhou2025onerec, liu2025llm}, e.g., three-level codes with size 4096.
However, in our findings, the naive Res-Kmeans method has a higher risk of codebook collision, where more than 30\% of Semantic IDs have multiple item candidates in the Shopping scenario.
We think the reason is that our system item environment follows a long-tail distribution, thus item representations are typically unevenly distributed in high-dimensional space (for example, there are dense and sparse regions). 
As a result, the codebook centroids learned by K-means will be located in the data-dense regions, resulting in an uneven distribution of the codebook in space since K-means only minimizes the average intra-cluster distance to centroids, without considering inter-cluster distances, as shown in Figure~\ref{fig:quant}(a).
Actually, we find that the first two-layer Res-Kmeans Code IDs are enough to identify an item's major category and usage; the latter quantization layer should reflect item-specific characteristics to reduce code collisions.
To balance fitting the real distribution and modeling item-specific characteristics, we introduce Finite Scalar Quantization (FSQ)~\cite{mentzer2023finite} to replace the last Res-Kmeans step. 
Unlike K-means, FSQ uses a predefined uniform quantization grid that is independent of the training item distribution, as shown in Figure~\ref{fig:quant}(b).
In this way, we can utilize the Res-Kmeans `layer-by-layer approximation' advantage to capture the core distribution, and use the FSQ `rule-based quantization' advantage to ensure fine-grained item semantics and reduce conflicts for more precise quantification.

Based on the reasoning item alignment and Res-KmeansFSQ quantization mechanisms, QARM V2 can generate accurate item LLM embeddings with appropriate business semantics to support GSU, while generalized and learnable Semantic IDs provide fine-grained interest modeling at the ESU side.
Overall, our contributions are as follows:
\begin{itemize}
    \item We present QARM V2 to produce more powerful LLM-based embeddings and Semantic IDs for user interest sequence modeling. Its embedding-enhanced GSU and Semantic-ID-enhanced ESU paradigm is verified in shopping, advertising, and live-streaming scenarios at Kuaishou.
    \item We propose a reasoning item alignment mechanism to reduce exposure noise and enhance underlying logical correlations in real-world business knowledge, alleviating the representation-unmatching problem. We also propose the Res-KmeansFSQ quantitative code mechanism to trade off between the unbalanced real-world long-tail distribution of Res-Kmeans and the balanced rule-based distribution of FSQ to avoid code conflicts.
    \item Our QARM V2 has been widely deployed on multiple services, supporting 400 million active users daily.
\end{itemize}

\begin{figure*}[t!]
  \centering
  \includegraphics[width=18cm,height=11.2cm]{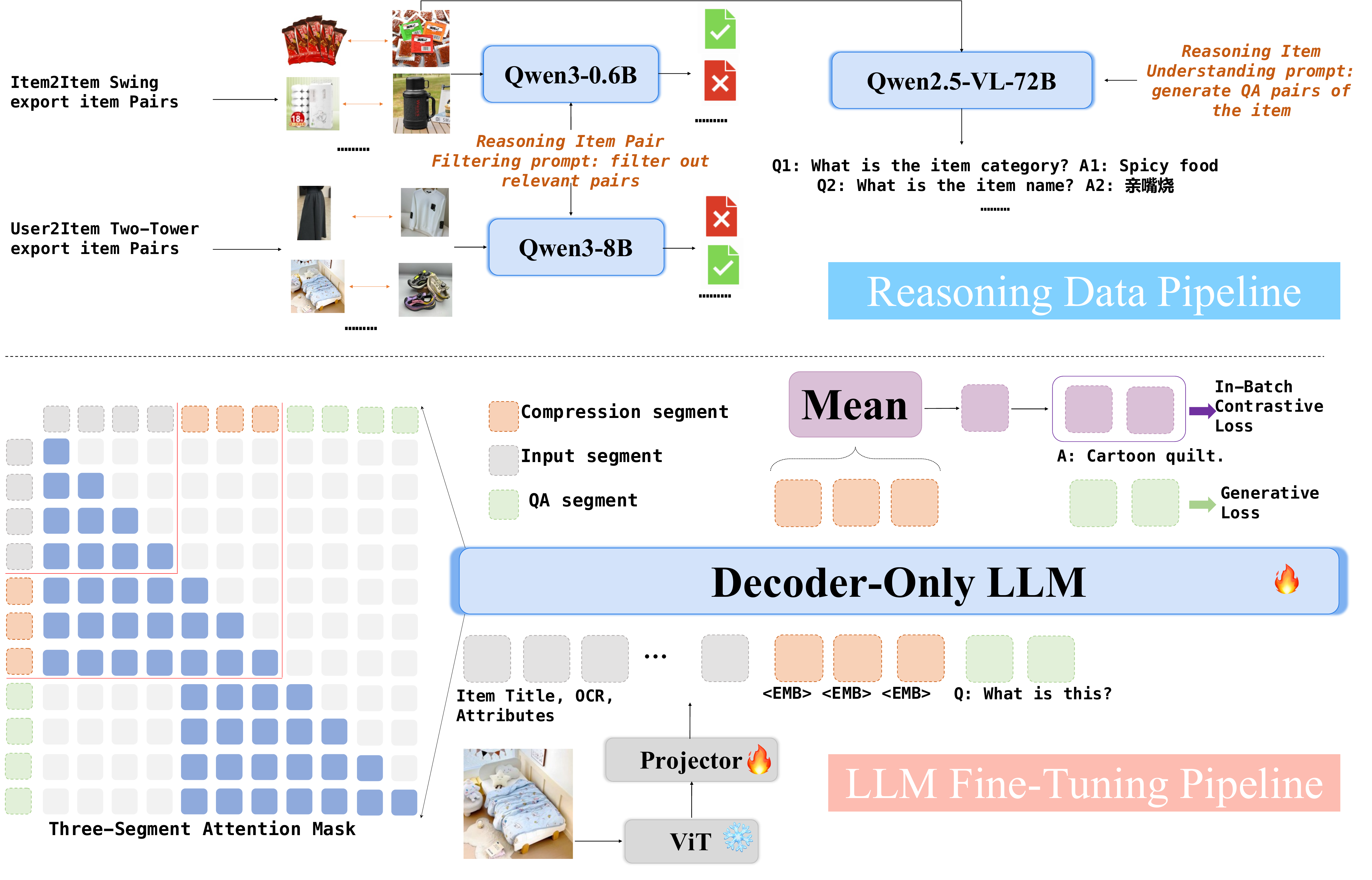}
    \caption{
    For the data pipeline, we first utilize the latest LLM to identify the item pairs relevance via their title, and then generating the corresponding question-answering pair based on each item's visual/text data. In LLM fine-tuning, we modify the attention mask to split the input token sequence as three segment, and then conduct contrastive and generation task to fine-tuning the LLM backbone to obtain the item semantic embedding with appropriate business knowledge.}
    \label{fig:mainarch}
\end{figure*}

\section{Methodology}
In this section, we introduce: (1) how the reasoning item alignment mechanism adapts the LLM to business knowledge; (2) how the Res-KmeansFSQ quantitative code mechanism produces the Semantic IDs; (3) how LLM embeddings enhance GSU and Semantic IDs enhance ESU to support users' interest sequence modeling.

\subsection{Reasoning Item Alignment}
Utilizing multimodal information to enhance RecSys performance is a long-term hot topic in both academic and industry communities.
Nowadays, the common consensus and experiments demonstrate that straightforwardly adding pre-trained multimodal embeddings as additional features brings very limited gains to RecSys, e.g., CLIP for image/frame compression~\cite{radford2021learning}, BERT for item ASR/OCR/title compression~\cite{radford2021learning}.
The reason is that early methods always utilize multiple `expert' pre-trained models (BERT,CLIP) to establish item relationships through visual-textual similarity, which can only identify low-level look-alike resemblance, failing to capture the underlying semantic differences between items.
Another reason is that there is a lack of efficient training objectives to bridge the gap between LLM and RecSys.
Previous approaches typically adopt contrastive learning between the image and text of the same item, or contrastive learning with the corresponding learned RecSys-ID embedding.
These methods are not only inefficient but also fail to generalize: our RecSys mission is not to recommend similar items our users have already seen, but rather to discover new items of interest expanded from their historical behaviors.
That is the desired role of LLMs in RecSys: to provide business-specific world knowledge signals that RecSys models are hard to capture.

In this section, we provide a detailed high-quality data construction pipeline, model backbone, and efficient training objectives to fulfill the gap between LLM and RecSys.

\subsubsection{Reasoning Supervision Data Pipeline}
In QARM V2, we have two categories of data: Item2Item contrastive data and item understanding data.
For the Item2Item contrastive data, we also choose to export them from our online serving retrieval models as the seed:
\begin{itemize}
    \item Exploitation Item2Item statistic retrieval model: directly export stable higher similarity item pairs from statistic Swing methods.
    \item Exploration User2Item Two-Tower retrieval model: for each positive item, from the recent positive items set select the most similar item in ID space.
\end{itemize}
In this way, we could collect a large-scale Item2Item data corpus; however, some of them are noisy supervision signals: the statistic Item2Item methods provide item pairs that are easily affected by hot-popular exposure bias, and the two-tower User2Item methods provide item pairs that are totally unrelated in terms of multi-modal content and underlying logic.
To reject low-quality noisy samples and filter more comprehensive and high-quality item pairs, we utilize the pre-trained reasoning model to judge whether item pairs are truly highly connected. The prompt is:
\begin{tcolorbox}[colback=gray!10!white,colframe=gray!50!black,title=Reasoning Prompt in Item Pair Filtering]
\small
Item-1 information: \#Title, \#Attributions, ...;\\
Item-2 information: \#Title, \#Attribution, ...;\\
Please analyze the potential correlation between two Items, such as semantic correlation (e.g., similar concepts or uses), complementary purchasing relationships (often purchased simultaneously or used together). Examples: \\
(1) (Fishing rod, parasol): Highly related usage scenarios; \\
(2) (Fishing rod, outdoor jacket): Shared outdoor activity scenario; \\
If the product pair correlation is strong, reply: `<answer>Yes</answer>`. Otherwise, reply: `<answer>No</answer>`.
\end{tcolorbox}
Specifically, we deploy the latest reasoning models~\cite{yang2025qwen3} to filter the item pair corpus, and only utilize the item's title and attributions as reasoning input. We use a smaller Qwen3-0.6B for exploitation Item2Item Swing item pairs and a larger Qwen3-8B for exploration User2Item two-tower item pairs to balance inference cost.
Overall, we reject 10\%+ Item2Item exported training samples and 70\%+ User2Item item pairs.

Instead of contrastive item-pair supervision, we also construct question-answering pairs for deeper item understanding, especially for complex short-video and live-streaming media.
To obtain high-quality question-answering pairs, we feed more fine-grained visual and text input to larger LLMs, such as Gemini~\cite{comanici2025gemini} and Qwen2.5-VL-72B~\cite{bai2025qwen2}.
The prompt is:
\begin{tcolorbox}[colback=gray!10!white,colframe=gray!50!black,title=Reasoning Prompt in Item Understanding]
\small
Item information: \#Title \#Attributes  \#Images, \#OCR, \#ASR, ...;\\
Please generate ten instruction questions as diverse as possible. These questions are about facts or an understanding and evaluation of relevant content. Do not ask any questions that cannot be answered confidently. \\
You need to return the result in the following form:\\
1. \{"Question":..., "Answer":...\}\\
2. \{"Question":..., "Answer":...\}
\end{tcolorbox}
By running this data generation pipeline, we could collect a large number of instruction question-answering pairs according to our provided prompt guidelines and item information.

\subsubsection{Decoder-Only GPT Embedding Generator}
As we know, the (current) SOTA LLMs all follow the decoder-only GPT paradigm; they aim to predict the next token according to the history token sequence, not a compressed summary embedding~\cite{achiam2023gpt,liu2024deepseek}.
As a result, there exists an unfriendly setting: it is better to provide a compressed embedding to represent the given visual-text token semantics in a downstream RecSys model.
How can we transform a decoder-only GPT paradigm LLM into a promising embedding generator?
To our knowledge, the most widely used method is to insert a special token as a single hint (e.g., \texttt{<EMB>}), and then use its last-layer hidden state as the `compressed embedding'.
Nevertheless, it severely modifies the generation-style NTP pre-training task into embedding compression, thus a question is raised: can we unify embedding compression and NTP at the same time?
In QARM V2, we give a three-segment trick to achieve the goal.
Specifically, we split the token input as:
\begin{itemize}
    \item \textit{Input Segment}: the given item's title, images, OCR and others.
    \item \textit{Compression Segment}: multiple special tokens \texttt{<EMB>} as compression units, which can attend to the input segment tokens only.
    \item \textit{QA Segment}: the last question-answer tokens, which can attend to the former segments only.
\end{itemize}
In this setting, the Input Segment can be seen as an information provider, the Compression Segment plays the role of compressing all inputs to generate embeddings, while the last QA Segment ensures that all fine-grained information can be extracted.
As shown in Figure~\ref{fig:mainarch}, the LLM can be optimized with the NTP generation task and embedding contrastive task at the same time.
In practice, we utilize the gradient cache technique to train the LLM in a large batch size environment, and a warm-start technique to anneal the output segment attention area (for the Output Segment, all Input/Query Segments can be seen at the training beginning, while reducing the Input Segment attention weight to zero step by step, for stable training).
Based on our contrastive item pair and generated item understanding data pipeline, and the three-segment LLM design, we can fine-tune the LLM as a stable embedding generator.
For the sake of better understanding, we do not show the formulation in this section; the overall workflow is shown in Figure~\ref{fig:mainarch}.

\subsection{Res-KmeansFSQ Quantitative Code}
Up to now, we have introduced how to fine-tune a pre-trained LLM to align with our business demands; next we show how it generates the Semantic IDs.
First of all, we denote the item embedding inference process as follows:
\begin{equation}
\small
\begin{split}
\mathbf{m} = \texttt{LLM}(\texttt{Item\_Text\_Vision\_Tokens}),
\end{split}
\label{iteminf}
\end{equation}
where the $\mathbf{m}\in \mathbb{R}^d$ denotes the item semantic embedding that will be used in the downstream RecSys model~\footnote{The LLM side embedding always has a large dimension, such as 3,000+.}.
%
%
In this paper, we focus on the QARM Res-Kmeans~\cite{luo2024qarm} quantification technique to enhance code ID quality and alleviate the code conflict problem (one Semantic ID mapping multiple real items).
In our system, the item distribution is always an uneven long-tail distribution (for example, there are dense and sparse regions), and the K-means optimization objective minimizes the average intra-cluster distance to centroids without considering inter-cluster distances, causing the naive three-level Res-Kmeans to show higher code conflict.

Actually, in our observation, we find that the first two-layer Res-Kmeans Code IDs are enough to identify an item's major category and usage.
This phenomenon motivates us to let the last layer quantization identify item-specific characteristics rather than model the real item distribution.
To this end, we introduce a hybrid quantization strategy that combines Res-Kmeans and Finite Scalar Quantization (FSQ). 
We retain Res-Kmeans for the first two layers to capture the coarse-grained item category and usage patterns, while replacing the last layer with FSQ to provide uniformly distributed codes for fine-grained item distinction.
This hybrid design allows us to leverage the advantages of both techniques: Res-Kmeans excels at learning data-adaptive centroids that reflect the natural clustering structure of item embeddings, while FSQ provides a deterministic and collision-free code assignment that effectively alleviates the code conflict problem in the tail region of the item distribution.
Particularly, the Res-Kmeans method has the following steps:
\begin{enumerate}[leftmargin=*,align=left]
\item \textit{Data collection}: Collecting a larger item embedding set $\mathbf{M}\in \mathbb{R}^{N\times d}$ sampled from the real distribution (e.g., $N>10,000,000$).
\item \textit{Initialized Two-layer Residual codebook training}: According to the training source $\mathbf{M}$, we first train the codebook with K-means as follows:
\begin{equation}
    \begin{split}
    \mathbf{C}^1 = \texttt{Kmeans}(\mathbf{M},& K), \quad \mathbf{M}^1 = \mathbf{M}-\texttt{NearestRep}(\mathbf{M}, \mathbf{C}^1)\\
    \mathbf{C}^2 &= \texttt{Kmeans}(\mathbf{M}^1, K)
    \end{split}
    \label{rqcodebooktrain}
\end{equation} 
where the \texttt{NearestRep} denotes the nearest K-means centroid representation, $K$ indicates the cluster hyper-parameter (e.g., 8192), and $\mathbf{C}^1, \mathbf{C}^2$ are the two-layer centroid representations.
\item \textit{Last-layer FSQ training}: According to the residual representation, we further train the last layer FSQ as follows:
\begin{equation}
    \begin{split}
    \mathbf{M}^2 &= \mathbf{M}^1 -  \texttt{NearestRep}(\mathbf{M}^1, \mathbf{C}^2) \\
    \mathbf{Z} &= \texttt{FSQ}(\mathbf{M}^2) = \lfloor L \cdot \sigma(\mathbf{M}^2\mathbf{W}) \rceil
    \end{split}
    \label{fsqtrain}
\end{equation} 
where $\mathbf{M}^2$ denotes the final residual after two-layer quantization, $\mathbf{W}\in \mathbb{R}^{d\times 13}$ is a learnable projection matrix, $\sigma(\cdot)$ is the sigmoid function that normalizes values to $[0,1]$, $L$ is the number of quantization levels per dimension (e.g., $L=2$), and $\lfloor \cdot \rceil$ denotes rounding to the nearest integer, and $\mathbf{Z}\in\mathbb{R}^n$ indicates the FSQ code.  
\end{enumerate}

\subsection{Semantic Enhanced GSU/ESU Workflow}
In this section, we dive into the RecSys model to describe how the generated Semantic embeddings and Semantic IDs are utilized in GSU and ESU enhancement.

\subsubsection{GSU Enhancement.}
In the GSU (General Selection Unit) stage, we directly leverage the LLM embedding $\mathbf{m}$ from Equation~\ref{iteminf} for efficient candidate retrieval. Specifically, we store all item LLM embeddings in a long-term embedding storage~\footnote{Typically, we perform PCA dimensionality reduction to avoid storage pressure.}. Given the target item representation $\mathbf{m}$ and a user's historical interaction sequence $\{i_1, i_2, \dots, i_T\}$ with corresponding LLM embeddings $\{\mathbf{m}_1, \mathbf{m}_2, \dots, \mathbf{m}_T\}$, we retrieve the top-$k$ most relevant historical items via inner product similarity:
\begin{equation}
    \{\mathbf{m}_1, \mathbf{m}_2, \dots, \mathbf{m}_k\} = \texttt{Top-K-GSU}(\{\mathbf{m}_1, \mathbf{m}_2, \dots, \mathbf{m}_T\}, \mathbf{m}, k)
\end{equation}
where the \texttt{Top-K GSU} operator selects the $k$ historical items with highest $\langle \mathbf{m}_i, \mathbf{m} \rangle$ scores. This semantic-aware retrieval captures deep item understanding from the LLM, enabling more effective subsequence selection compared to ID-based or rule-based methods.

\subsubsection{ESU Enhancement.}
In the ESU (Exact Selection Unit) stage, we utilize the multi-level Semantic IDs (SIDs) obtained from ResKmeansFSQ as learnable discrete features. Each item is represented by its original ItemID together with the three-level SIDs $\{c^1, c^2, c^3\}$, which are mapped to learnable embeddings via lookup tables. The ESU then performs target-attention over the retrieved subsequence:
\begin{equation}
\small
\begin{split}
\texttt{ESU} = \texttt{Target}&\texttt{Attention}((\mathbf{I}, \textbf{c}^1, \textbf{c}^2, \textbf{c}^3),\\
&\{(\mathbf{I}_1, \textbf{c}^1_1, \textbf{c}^2_1, \textbf{c}^3_1), \dots, (\mathbf{I}_k, \textbf{c}^1_k, \textbf{c}^2_k, \textbf{c}^3_k)\}),\\
\hat{y}^\texttt{ctr}, \hat{y}^\texttt{cvr},& \dots \leftarrow \texttt{MoE}(\texttt{ESU}, \texttt{Others})
\end{split}
\label{esumodeling}
\end{equation}
where the target item features $(\mathbf{I}, \textbf{c}^1, \textbf{c}^2, \textbf{c}^3)$ serve as the query, and the historical subsequence features serve as keys and values.
The attended representation is then fed into a multi-task MoE module for CTR/CVR prediction~\cite{cheng2025choruscvr}. The entire model is optimized end-to-end with the multi-task binary cross-entropy loss:
\begin{equation}
\mathcal{L} = - \sum_{\texttt{xtr}\in\{\texttt{ctr}, \texttt{cvr}, \dots\}} \big(y^{\texttt{xtr}}\log{(\hat{y}^{\texttt{xtr}})} + (1-y^{\texttt{xtr}})\log{(1-\hat{y}^{\texttt{xtr}})}\big)
\label{esuloss}
\end{equation}
Through this end-to-end optimization, the SID embeddings are jointly learned with the ranking model, allowing the hierarchical semantic structure encoded in SIDs to directly benefit the downstream prediction tasks.

\begin{table}[t!]
\centering
\caption{Performance comparison (\%) at Amazon Book.}
\setlength{\tabcolsep}{3.5pt}{

\begin{tabular}{c|cccc}
\toprule
Baselines & DIN & SIM-hard & SIM-soft & QARM V2 \\
\midrule
AUC & 67.69 & 67.15 & 69.57 & 70.33 \\
\bottomrule
\end{tabular}
}
\label{amazondata}
\end{table}

\section{Experiments}
To valid our reasoning item alignment and ResKmeansFSQ effectiveness, we conduct extensive experiments in Advertising, Shopping and Live-streaming services, and Amazon public dataset.

\begin{table*}[t!]
\centering
\caption{Offline results(\%) in terms of AUC, UAUC and GAUC on Advertising/Shopping\#1 at Kuaishou.}
\setlength{\tabcolsep}{6pt}{
\begin{tabular}{l|ccc|ccccccccc}
\toprule
\multirow{4}{*}{\makecell{Model\\Variants}} 
& \multicolumn{3}{c}{Advertising} & \multicolumn{9}{c}{Shopping\#1}   \\ 
\cmidrule(r){2-4} \cmidrule(r){5-13} & \multicolumn{3}{c}{CTCVR} & \multicolumn{3}{c}{CTR}  & \multicolumn{3}{c}{CVR} & \multicolumn{3}{c}{CTCVR}  \\ 
\cmidrule(r){2-4} \cmidrule(r){5-7} \cmidrule(r){8-10} \cmidrule(r){11-13} & AUC & UAUC & GAUC & AUC & UAUC & GAUC & AUC & UAUC & GAUC & AUC & UAUC & GAUC\\
\hline
Baseline Model & 87.02 & 62.98 & 63.06 & 81.70 & 70.67 & 71.45 & 89.52 & 71.61 & 72.50 & 90.39 & 74.99 & 74.97 \\
+QARM V2 & 87.43 & 63.97 & 64.16 & 81.79 & 70.85 & 71.63 & 89.69 & 71.83 & 72.76 & 90.59 & 75.34 & 75.37 \\
\bottomrule
\end{tabular}
}
\label{offline_adv_shop1}
\end{table*}

\begin{table*}[t!]
\centering
\caption{Offline results(\%) in terms of AUC, UAUC and GAUC on Shopping services at Kuaishou.}
\setlength{\tabcolsep}{6pt}{
\begin{tabular}{l|cccccc|cccccc}
\toprule
\multirow{4}{*}{\makecell{Model\\Variants}} 
& \multicolumn{6}{c}{Shopping\#2} & \multicolumn{6}{c}{Shopping\#3}   \\ 
\cmidrule(r){2-7} \cmidrule(r){8-13} & \multicolumn{3}{c}{CTR} & \multicolumn{3}{c}{CVR}  & \multicolumn{3}{c}{CTR} & \multicolumn{3}{c}{CVR}  \\ 
\cmidrule(r){2-4} \cmidrule(r){5-7} \cmidrule(r){8-10} \cmidrule(r){11-13} & AUC & UAUC & WUAUC & AUC & UAUC & GAUC & AUC & UAUC & GAUC & AUC & UAUC & GAUC\\
\hline
Baseline Model & 85.95 &65.75 & 65.97 & 87.29 & 69.5 & 69.18 & 86.77 & 64.96 & 66.41 & 89.41 & 65.38 & 67.54 \\
+QARM V2 & 86.10 & 65.94 & 66.18 & 87.38 & 69.58 & 69.34 & 86.82 & 65.08 & 66.53 & 89.46 & 65.44 & 67.59 \\
\bottomrule
\end{tabular}
}
\label{offline_shop23}
\end{table*}

\begin{table*}[t!]
\centering
\caption{Offline results(\%) in terms of AUC, UAUC and GAUC on Live-streaming services at Kuaishou.}
\setlength{\tabcolsep}{6pt}{
\begin{tabular}{l|cccccccccccc}
\toprule
\multirow{3}{*}{\makecell{Model\\Variants}} 
& \multicolumn{12}{c}{Live-streaming} \\ 
\cmidrule(r){2-13} 
& \multicolumn{3}{c}{Click} & \multicolumn{3}{c}{Gift} & \multicolumn{3}{c}{Long View} & \multicolumn{3}{c}{Follow} \\ 
\cmidrule(r){2-4} \cmidrule(r){5-7} \cmidrule(r){8-10} \cmidrule(r){11-13} 
& AUC & UAUC & GAUC & AUC & UAUC & GAUC & AUC & UAUC & GAUC & AUC & UAUC & GAUC \\
\midrule
Baseline Model & 82.68 & 63.48 & 63.61 & 97.65 & 70.02 & 70.33 & 83.70 & 67.28 & 67.26 & 83.70 & 66.83 & 66.82 \\
+QARM V2 & 82.78 & 63.60 & 63.87 & 97.69 & 70.11 & 70.55 & 83.80 & 67.60 & 67.76 & 83.76 & 66.94 & 67.10 \\
\bottomrule
\end{tabular}
}
\label{livestreaming_results}
\end{table*}

\subsection{Offline Performance}

\textbf{Academic Datasets.}
To validate the generalization of QARM V2, we first conduct experiments on the public Amazon dataset.
Specifically, we preprocess the data as follows: (1) samples with rating $\geq$ 4 are labeled as positive, otherwise negative; (2) we maintain users with sequence length $\geq$ 20; each user contributes one positive and one negative sample, and 15\% of users are randomly split into the test set.
In model training, we retrieve top-50 items for the last item to support ESU calculation.
As shown in Table~\ref{amazondata}, QARM V2 achieves the best AUC of 70.33\%, outperforming all baselines including DIN (67.69\%, using users' latest interaction items), SIM hard (67.15\%, searching items with the same tag), and SIM soft (69.57\%, searching top-$k$ related items by learnable item ID embedding). This demonstrates that our reasoning-based item alignment and Res-KmeansFSQ quantization can effectively enhance GSU search accuracy and ESU end-to-end optimization.

\textbf{Industry Datasets.}
We further evaluate QARM V2 on large-scale industrial datasets from Kuaishou, covering Advertising, Shopping, and Live-streaming services. 
Empirically, AUC/UAUC/GAUC are the vital offline metrics for evaluating model performance; on average, about 0.1\% improvement is enough to contribute business gains.
As shown in Table~\ref{offline_adv_shop1}, Table~\ref{offline_shop23}, and Table~\ref{livestreaming_results}, QARM V2 consistently improves over the baseline across all scenarios and metrics. In Advertising, GAUC improves by +1.10\% (63.06\%$\rightarrow$64.16\%). In Shopping\#1, the CTCVR GAUC gains +0.40\% (74.97\%$\rightarrow$75.37\%). In Live-streaming, Long View GAUC achieves the largest improvement of +0.50\% (67.26\%$\rightarrow$67.76\%). These results confirm the effectiveness of QARM V2 in enhancing both GSU retrieval and ESU ranking across diverse industrial recommendation scenarios.
A valuable but not directly observable phenomenon is that all our experiments are conducted without sequence deduplication, as we observe that deduplicated sequences consistently yield lower offline AUC than non-deduplicated ones. For example, in Live-streaming, out of top-100 interactions, there are only 23 unique authors on average.
The non-deduplicated sequence captures users' repeated engagement patterns on the same author, such as viewing duration variations and time-of-day preferences, which provide valuable signals for preference modeling.

\begin{table}[t!]
\centering
\caption{Online A/B testing at Advertising Scenario.}
\setlength{\tabcolsep}{3.5pt}{

\begin{tabular}{c|ccc}
\toprule
\multirow{2}{*}{\makecell{Scenarios}}                  & \multicolumn{3}{c}{Advertising Metrics}                                                                             \\
\cmidrule(r){2-4}  
    & Exposure &  Cost   & Revenue \\
\midrule
Advertising & +1.321\% & +3.942\% & +4.873\% \\
\bottomrule
\end{tabular}
}
\label{mainonlineadv1}
\end{table}

\subsection{Online Performance}
We deployed QARM V2 in production across multiple business scenarios and conducted multi-week online A/B testing.
As reported in Table~\ref{mainonlineadv1}, Table~\ref{mainonlineshop}, and Table~\ref{tab:detailed_results_optimized}~\footnote{Our experiments conduct at main-traffic Ads/shop/live-streaming scenarios.}, our QARM V2 consistently delivered significant online gains across our main-stream scenarios.
In Advertising services, Revenue increased by +4.873\% with Cost improved by +3.942\%.
In Shopping services, Shopping\#2 achieved the most substantial gains with GMV +5.612\% and Order +4.834\%, while Shopping\#1 and Shopping\#3 also demonstrated consistent improvements across all metrics.
These results validate that our reasoning-based item alignment and Res-KmeansFSQ quantization effectively enhance both GSU retrieval quality and ESU ranking.

\subsection{Reasoning Item Alignment Analysis}
To evaluate the GSU retrieval capability of our reasoning-based LLM embeddings, we deploy an item-to-item retrieval service that uses users' real-time clicks and orders as ground truth. Specifically, for each user, we use the last 10 historically clicked trigger items to retrieve 50 candidates per item trigger (totaling up to 500 candidates), and compute the hit rate against the user's actual interactions (clicks and orders).
As shown in Table~\ref{iaable}, QARM-V2 achieves substantial improvements over QARM across all metrics. For click prediction, HR@200 improves from 7.77\% to 12.5\% (+60.9\% relative), and HR@500 improves from 9.7\% to 15.4\% (+58.8\% relative). For order prediction, HR@200 improves from 11.3\% to 20.0\% (+77.0\% relative), and HR@500 improves from 20.0\% to 23.0\%. Notably, the significant gains at HR@200 indicate that QARM-V2 produces more reasonable similarity scores, ranking truly relevant items higher rather than requiring deeper retrieval to achieve hits. These results validate that our reasoning-based item alignment could filter the noise item pairs to generate LLM embeddings that better capture user behavioral preferences.

\begin{table}[t!]
\centering
\caption{Online A/B testing results at Shopping services.}
\setlength{\tabcolsep}{7pt}{

\begin{tabular}{c|ccc}
\toprule
\multirow{2}{*}{Scenarios}  & \multicolumn{3}{c}{Shopping Metrics}                                                                             \\
\cmidrule(r){2-4}  
    & GMV &  Order   & Exposure \\
\midrule
Shopping\#1 & +3.208\% & +1.919\% & +0.427\% \\
\midrule
\multirow{1}{*}{\makecell{Shopping\#2}}  & +5.612\% &  +4.834\% & +3.739\% \\

\midrule
\multirow{1}{*}{\makecell{Shopping\#3}}  &+1.037\% & +1.221\%& +1.487\% \\

\bottomrule
\end{tabular}
}
\label{mainonlineshop}
\end{table}

\begin{table*}[t]
\centering
\caption{Online A/B testing results at Live-streaming services, at the largest live-stream traffic.}
\label{tab:detailed_results_optimized}
\setlength{\tabcolsep}{4pt} 
\begin{tabular}{@{}l|l|cccc|ccc@{}}
\toprule
\multirow{2}{*}{Scenarios} & \multirow{2}{*}{Item Group} & \multicolumn{4}{c}{Core Metrics} & \multicolumn{3}{c}{Interact Metrics} \\
\cmidrule(r){3-6} \cmidrule(r){7-9}
& & Click & Watch Time & Watch Count & Gift Count & Like & Comment & Follow \\
\midrule
\multirow{2}{*}{Live-streaming\#1}
& Cold-start & +3.231\% & +2.961\% & +1.609\% & +1.807\% & \makecell{--} & \makecell{--} & \makecell{--} \\
& Others & +0.611\% & +0.753\% & +0.489\% & +2.917\% & +3.215\% & +1.184\% & +1.158\% \\
\multirow{2}{*}{Live-streaming\#2}
& Cold-start & +0.890\% & +0.998\% & +0.998\% & +0.918\% & \makecell{--} & \makecell{--} & \makecell{--} \\
& Others & +0.304\% & +0.482\% & +0.434\% & +2.917\% & +0.464\% & +0.826\% & +0.418\% \\
\bottomrule
\end{tabular}
\end{table*}

\begin{table}[t!]
\centering
\caption{Reasoning Item Alignment Improvement.}
\setlength{\tabcolsep}{5pt}{

\begin{tabular}{l|cc|ccc}
\toprule
\multirow{2}{*}{\makecell{Method}} & \multicolumn{2}{c|}{Click} & \multicolumn{2}{c}{Order} \\
\cmidrule(r){2-3} \cmidrule(r){4-5}
&HR@200 & HR@500 &  HR@200   & HR@500  \\
\midrule
QARM &7.77\% &9.7\% &11.3\% &13.0\%  \\
QARM V2 &12.5\% &15.4\% &20.0\% &23.0\% \\
\bottomrule
\end{tabular}
}
\label{iaable}
\end{table}

\begin{table}[t!]
\centering
\caption{Code Conflict Component Analysis.}
\setlength{\tabcolsep}{2pt}{

\begin{tabular}{l|cc|ccc}
\toprule
Methods &Collision & EdgeNum &  HR@1   & HR@10  \\
\midrule
QARM ResKmeans &77.92\% &129.33 &80.3\% &97.79\%  \\
QARM V2 ResKmeans &52.25\% &8.41 &91.9\% &99.75\%  \\
QARM V2 ResKmeansFSQ &32.39\% &2.5 &95.2\% &99.9\%  \\
\bottomrule
\end{tabular}
}
\label{codeconflict}
\vspace{-0.4cm}
\end{table}

\subsection{Res-KmeansFSQ Code Conflict Analysis}
To evaluate the effectiveness of our Res-KmeansFSQ in reducing code conflicts, we deploy a KGNN (a graph edge storage engine at Kuaishou) service that hashes SID sequences into hash IDs and maintains (hash\_id, item\_id) edges from the online sample stream.
For QARM V2-Res-Kmeans training, the MSE loss (per element) is about 0.37 at the first level, 0.26 at the second level, and 0.20 at the third level (FSQ training loss is much higher than K-means).
Based on these, we first obtain the SID for a given query item, then perform reverse lookup in KGNN to retrieve candidate item IDs. We measure: (1) \textit{Collision}: the percentage of items sharing the same SID with others; (2) \textit{EdgeNum}: the average number of item IDs returned per SID query; (3) \textit{HR@K}: the Hit Rate of retrieving the correct item within top-K results.
As shown in Table~\ref{codeconflict}, higher collision rates lead to more items sharing the same SID, making it harder to retrieve the correct item, thus requiring larger K to ensure a hit. 
Specifically, compared to QARM-ResKmeans (3 * 4096), our QARM-V2-ResKmeans (3 * 4096) reduces the collision rate from 77.92\% to 52.25\% and EdgeNum from 129.33\% to 8.41\%, improving HR@1 from 80.3\% to 91.9\%, verifying our reasoning item alignment mechanism. By further incorporating FSQ at the last layer, QARM-V2-ResKmeansFSQ (3 * 4096) achieves the lowest collision rate of 32.39 and EdgeNum of only 2.5, with HR@1 reaching 95.2\%. These results validate that our hybrid quantization strategy effectively alleviates code conflicts, enabling precise item retrieval.

\subsection{GSU Case Analysis}
\begin{figure}[t!]
  \centering
  \includegraphics[width=8cm,height=3cm]{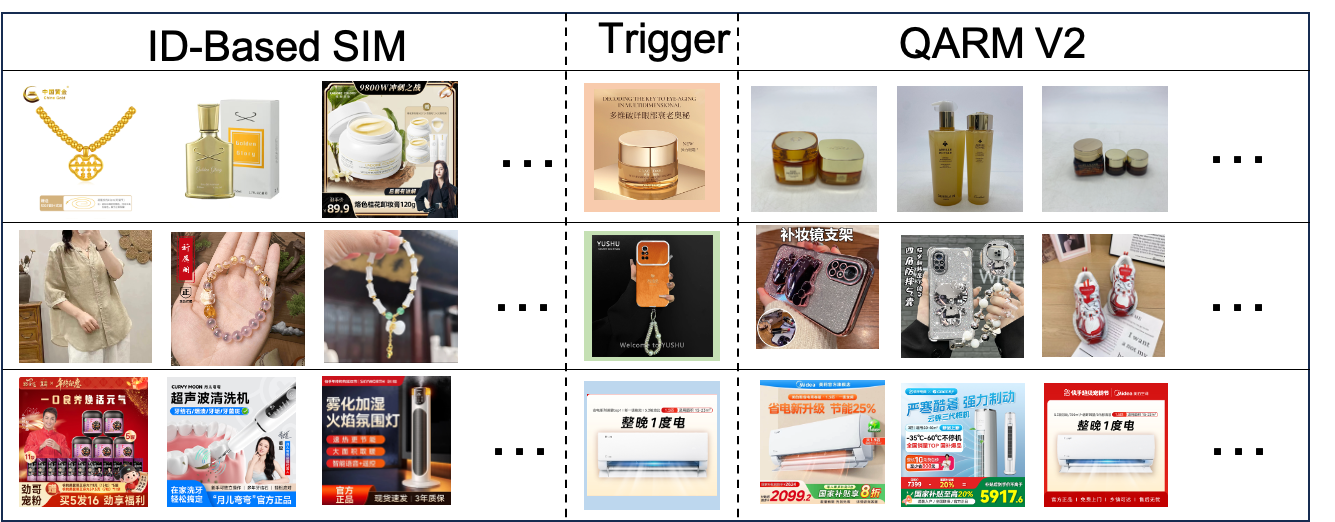}
    \caption{Visualization of exclusive retrieved items.}
    \label{fig:gsucase}
\end{figure}

To intuitively evaluate the retrieval quality within the General Search Unit (GSU) stage, Table~\ref{exclusive} and Figure~\ref{fig:gsucase} present a comparative visualization of retrieval results between our proposed QARM-V2 and the baseline SIM model. 
The exclusive retrieval rate measures items that are uniquely recalled by QARM-V2 but not by the baseline GSU. We report this metric in Table~\ref{exclusive} to quantify the ability to expand candidate sets beyond repeated exposure.
These exclusive items are often semantically related yet visually dissimilar, indicating that QARM-V2 can expand users' interests beyond repeated exposure and discover novel but relevant candidates.
In Figure~\ref{fig:gsucase}, we show some cases of trigger items across diverse categories.
As observed in the figure, the baseline End2End ID-based SIM frequently retrieves "hard negative" items that are irrelevant to the trigger item in both category and semantics (e.g., retrieving unrelated jewelry for a phone accessory target).
In contrast, QARM-V2 demonstrates superior semantic alignment and robustness. Across all presented cases, the candidate sets retrieved by our method maintain high categorical consistency and semantic relevance to the target items. Whether in scenarios requiring visual discrimination or fine-grained semantic understanding, QARM-V2 accurately identifies and retrieves reasonable items that align with the user's underlying interests. These results confirm that QARM-V2 significantly enhances the relevance of user sequence modeling.

\begin{table}[t!]
\centering
\caption{QARM V2 GSU exclusive retrieve rate.}
\setlength{\tabcolsep}{2pt}{

\begin{tabular}{l|ccc}
\toprule
Sequence &Click(Top50) & Order(Top30) &  Exposure(Top30)  \\
\midrule
Exclusive Rate & 63.6\% &57.9\% &65.1\% \\
\bottomrule
\end{tabular}
}
\label{exclusive}
\vspace{-0.4cm}
\end{table}

\section{Conclusions}
In this paper, we present QARM V2, a unified framework that leverages LLM semantic understanding to enhance lifelong user sequence modeling in industrial RecSys. 
Our approach introduces reasoning-based item alignment from both Item-to-Item and User-to-Item perspectives, generating business-aligned LLM embeddings for GSU retrieval. 
Further, we provide a three-segment LLM fine-tuning paradigm to transform a decoder-only LLM into an embedding generator with contrastive and generation tasks simultaneously.
For ESU, we propose Res-KmeansFSQ, a hybrid quantization method that combines Residual K-means with Finite Scalar Quantization, producing multi-level Semantic IDs with significantly lower code collision rates. 
Extensive experiments on Kuaishou's short-video and live-streaming platforms demonstrate that QARM V2 achieves significant improvements in both GSU and ESU capabilities, delivering substantial gains across multiple business scenarios.
Moreover, we conduct detailed ablation studies to highlight the LLM reasoning reject sampling is necessary to keep business distribution with world knowledge.
In future work, we will explore finer-grained, longer parallel VQ codes, which can be diffused to the original modality.
\newpage

\balance
\bibliographystyle{ACM-Reference-Format}
\bibliography{sample-base-extend.bib}


\begin{thebibliography}{28}


\ifx \showCODEN    \undefined \def \showCODEN     #1{\unskip}     \fi
\ifx \showISBNx    \undefined \def \showISBNx     #1{\unskip}     \fi
\ifx \showISBNxiii \undefined \def \showISBNxiii  #1{\unskip}     \fi
\ifx \showISSN     \undefined \def \showISSN      #1{\unskip}     \fi
\ifx \showLCCN     \undefined \def \showLCCN      #1{\unskip}     \fi
\ifx \shownote     \undefined \def \shownote      #1{#1}          \fi
\ifx \showarticletitle \undefined \def \showarticletitle #1{#1}   \fi
\ifx \showURL      \undefined \def \showURL       {\relax}        \fi
\providecommand\bibfield[2]{#2}
\providecommand\bibinfo[2]{#2}
\providecommand\natexlab[1]{#1}
\providecommand\showeprint[2][]{arXiv:#2}

\bibitem[Achiam et~al\mbox{.}(2023)]%
        {achiam2023gpt}
\bibfield{author}{\bibinfo{person}{Josh Achiam}, \bibinfo{person}{Steven Adler}, \bibinfo{person}{Sandhini Agarwal}, \bibinfo{person}{Lama Ahmad}, \bibinfo{person}{Ilge Akkaya}, \bibinfo{person}{Florencia~Leoni Aleman}, \bibinfo{person}{Diogo Almeida}, \bibinfo{person}{Janko Altenschmidt}, \bibinfo{person}{Sam Altman}, \bibinfo{person}{Shyamal Anadkat}, {et~al\mbox{.}}} \bibinfo{year}{2023}\natexlab{}.
\newblock \showarticletitle{Gpt-4 technical report}.
\newblock \bibinfo{journal}{\emph{arXiv}} (\bibinfo{year}{2023}).
\newblock


\bibitem[Bai et~al\mbox{.}(2023)]%
        {bai2023qwen}
\bibfield{author}{\bibinfo{person}{Jinze Bai}, \bibinfo{person}{Shuai Bai}, \bibinfo{person}{Yunfei Chu}, \bibinfo{person}{Zeyu Cui}, \bibinfo{person}{Kai Dang}, \bibinfo{person}{Xiaodong Deng}, \bibinfo{person}{Yang Fan}, \bibinfo{person}{Wenbin Ge}, \bibinfo{person}{Yu Han}, \bibinfo{person}{Fei Huang}, {et~al\mbox{.}}} \bibinfo{year}{2023}\natexlab{}.
\newblock \showarticletitle{Qwen technical report}.
\newblock \bibinfo{journal}{\emph{arXiv preprint arXiv:2309.16609}} (\bibinfo{year}{2023}).
\newblock


\bibitem[Bai et~al\mbox{.}(2025)]%
        {bai2025qwen2}
\bibfield{author}{\bibinfo{person}{Shuai Bai}, \bibinfo{person}{Keqin Chen}, \bibinfo{person}{Xuejing Liu}, \bibinfo{person}{Jialin Wang}, \bibinfo{person}{Wenbin Ge}, \bibinfo{person}{Sibo Song}, \bibinfo{person}{Kai Dang}, \bibinfo{person}{Peng Wang}, \bibinfo{person}{Shijie Wang}, \bibinfo{person}{Jun Tang}, {et~al\mbox{.}}} \bibinfo{year}{2025}\natexlab{}.
\newblock \showarticletitle{Qwen2. 5-vl technical report}.
\newblock \bibinfo{journal}{\emph{arXiv preprint arXiv:2502.13923}} (\bibinfo{year}{2025}).
\newblock


\bibitem[Chai et~al\mbox{.}(2025)]%
        {chai2025longer}
\bibfield{author}{\bibinfo{person}{Zheng Chai}, \bibinfo{person}{Qin Ren}, \bibinfo{person}{Xijun Xiao}, \bibinfo{person}{Huizhi Yang}, \bibinfo{person}{Bo Han}, \bibinfo{person}{Sijun Zhang}, \bibinfo{person}{Di Chen}, \bibinfo{person}{Hui Lu}, \bibinfo{person}{Wenlin Zhao}, \bibinfo{person}{Lele Yu}, {et~al\mbox{.}}} \bibinfo{year}{2025}\natexlab{}.
\newblock \showarticletitle{Longer: Scaling up long sequence modeling in industrial recommenders}. In \bibinfo{booktitle}{\emph{Proceedings of the Nineteenth ACM Conference on Recommender Systems}}. \bibinfo{pages}{247--256}.
\newblock


\bibitem[Chang et~al\mbox{.}(2023)]%
        {chang2023twin}
\bibfield{author}{\bibinfo{person}{Jianxin Chang}, \bibinfo{person}{Chenbin Zhang}, \bibinfo{person}{Zhiyi Fu}, \bibinfo{person}{Xiaoxue Zang}, \bibinfo{person}{Lin Guan}, \bibinfo{person}{Jing Lu}, \bibinfo{person}{Yiqun Hui}, \bibinfo{person}{Dewei Leng}, \bibinfo{person}{Yanan Niu}, \bibinfo{person}{Yang Song}, {et~al\mbox{.}}} \bibinfo{year}{2023}\natexlab{}.
\newblock \showarticletitle{TWIN: TWo-stage interest network for lifelong user behavior modeling in CTR prediction at kuaishou}. In \bibinfo{booktitle}{\emph{Proceedings of the 29th ACM SIGKDD Conference on Knowledge Discovery and Data Mining}}. \bibinfo{pages}{3785--3794}.
\newblock


\bibitem[Cheng et~al\mbox{.}(2025)]%
        {cheng2025choruscvr}
\bibfield{author}{\bibinfo{person}{Wei Cheng}, \bibinfo{person}{Yucheng Lu}, \bibinfo{person}{Boyang Xia}, \bibinfo{person}{Jiangxia Cao}, \bibinfo{person}{Kuan Xu}, \bibinfo{person}{Mingxing Wen}, \bibinfo{person}{Wei Jiang}, \bibinfo{person}{Jiaming Zhang}, \bibinfo{person}{Zhaojie Liu}, \bibinfo{person}{Liyin Hong}, {et~al\mbox{.}}} \bibinfo{year}{2025}\natexlab{}.
\newblock \showarticletitle{ChorusCVR: Chorus Supervision for Entire Space Post-Click Conversion Rate Modeling}.
\newblock \bibinfo{journal}{\emph{arXiv preprint arXiv:2502.08277}} (\bibinfo{year}{2025}).
\newblock


\bibitem[Comanici et~al\mbox{.}(2025)]%
        {comanici2025gemini}
\bibfield{author}{\bibinfo{person}{Gheorghe Comanici}, \bibinfo{person}{Eric Bieber}, \bibinfo{person}{Mike Schaekermann}, \bibinfo{person}{Ice Pasupat}, \bibinfo{person}{Noveen Sachdeva}, \bibinfo{person}{Inderjit Dhillon}, \bibinfo{person}{Marcel Blistein}, \bibinfo{person}{Ori Ram}, \bibinfo{person}{Dan Zhang}, \bibinfo{person}{Evan Rosen}, {et~al\mbox{.}}} \bibinfo{year}{2025}\natexlab{}.
\newblock \showarticletitle{Gemini 2.5: Pushing the frontier with advanced reasoning, multimodality, long context, and next generation agentic capabilities}.
\newblock \bibinfo{journal}{\emph{arXiv preprint arXiv:2507.06261}} (\bibinfo{year}{2025}).
\newblock


\bibitem[Covington et~al\mbox{.}(2016)]%
        {youtubednn}
\bibfield{author}{\bibinfo{person}{Paul Covington}, \bibinfo{person}{Jay Adams}, {and} \bibinfo{person}{Emre Sargin}.} \bibinfo{year}{2016}\natexlab{}.
\newblock \showarticletitle{Deep neural networks for youtube recommendations}. In \bibinfo{booktitle}{\emph{ACM Conference on Recommender Systems (RecSys)}}.
\newblock


\bibitem[Deng et~al\mbox{.}(2025)]%
        {deng2025onerec}
\bibfield{author}{\bibinfo{person}{Jiaxin Deng}, \bibinfo{person}{Shiyao Wang}, \bibinfo{person}{Kuo Cai}, \bibinfo{person}{Lejian Ren}, \bibinfo{person}{Qigen Hu}, \bibinfo{person}{Weifeng Ding}, \bibinfo{person}{Qiang Luo}, {and} \bibinfo{person}{Guorui Zhou}.} \bibinfo{year}{2025}\natexlab{}.
\newblock \showarticletitle{Onerec: Unifying retrieve and rank with generative recommender and iterative preference alignment}.
\newblock \bibinfo{journal}{\emph{arXiv preprint arXiv:2502.18965}} (\bibinfo{year}{2025}).
\newblock


\bibitem[Kang and McAuley(2018)]%
        {sasrec}
\bibfield{author}{\bibinfo{person}{Wang-Cheng Kang} {and} \bibinfo{person}{Julian McAuley}.} \bibinfo{year}{2018}\natexlab{}.
\newblock \showarticletitle{Self-attentive sequential recommendation}. In \bibinfo{booktitle}{\emph{IEEE international conference on data mining (ICDM)}}.
\newblock


\bibitem[Liu et~al\mbox{.}(2024b)]%
        {liu2024deepseek}
\bibfield{author}{\bibinfo{person}{Aixin Liu}, \bibinfo{person}{Bei Feng}, \bibinfo{person}{Bing Xue}, \bibinfo{person}{Bingxuan Wang}, \bibinfo{person}{Bochao Wu}, \bibinfo{person}{Chengda Lu}, \bibinfo{person}{Chenggang Zhao}, \bibinfo{person}{Chengqi Deng}, \bibinfo{person}{Chenyu Zhang}, \bibinfo{person}{Chong Ruan}, {et~al\mbox{.}}} \bibinfo{year}{2024}\natexlab{b}.
\newblock \showarticletitle{Deepseek-v3 technical report}.
\newblock \bibinfo{journal}{\emph{arXiv preprint arXiv:2412.19437}} (\bibinfo{year}{2024}).
\newblock


\bibitem[Liu et~al\mbox{.}(2024a)]%
        {kuaiformer}
\bibfield{author}{\bibinfo{person}{Chi Liu}, \bibinfo{person}{Jiangxia Cao}, \bibinfo{person}{Rui Huang}, \bibinfo{person}{Kai Zheng}, \bibinfo{person}{Qiang Luo}, \bibinfo{person}{Kun Gai}, {and} \bibinfo{person}{Guorui Zhou}.} \bibinfo{year}{2024}\natexlab{a}.
\newblock \showarticletitle{KuaiFormer: Transformer-Based Retrieval at Kuaishou}.
\newblock  (\bibinfo{year}{2024}).
\newblock


\bibitem[Liu et~al\mbox{.}(2025)]%
        {liu2025llm}
\bibfield{author}{\bibinfo{person}{Yueyang Liu}, \bibinfo{person}{Jiangxia Cao}, \bibinfo{person}{Shen Wang}, \bibinfo{person}{Shuang Wen}, \bibinfo{person}{Xiang Chen}, \bibinfo{person}{Xiangyu Wu}, \bibinfo{person}{Shuang Yang}, \bibinfo{person}{Zhaojie Liu}, \bibinfo{person}{Kun Gai}, {and} \bibinfo{person}{Guorui Zhou}.} \bibinfo{year}{2025}\natexlab{}.
\newblock \showarticletitle{LLM-Alignment Live-Streaming Recommendation}.
\newblock \bibinfo{journal}{\emph{arXiv preprint arXiv:2504.05217}} (\bibinfo{year}{2025}).
\newblock


\bibitem[Luo et~al\mbox{.}(2024)]%
        {luo2024qarm}
\bibfield{author}{\bibinfo{person}{Xinchen Luo}, \bibinfo{person}{Jiangxia Cao}, \bibinfo{person}{Tianyu Sun}, \bibinfo{person}{Jinkai Yu}, \bibinfo{person}{Rui Huang}, \bibinfo{person}{Wei Yuan}, \bibinfo{person}{Hezheng Lin}, \bibinfo{person}{Yichen Zheng}, \bibinfo{person}{Shiyao Wang}, \bibinfo{person}{Qigen Hu}, {et~al\mbox{.}}} \bibinfo{year}{2024}\natexlab{}.
\newblock \showarticletitle{QARM: Quantitative Alignment Multi-Modal Recommendation at Kuaishou}.
\newblock \bibinfo{journal}{\emph{arXiv}} (\bibinfo{year}{2024}).
\newblock


\bibitem[Mentzer et~al\mbox{.}(2023)]%
        {mentzer2023finite}
\bibfield{author}{\bibinfo{person}{Fabian Mentzer}, \bibinfo{person}{David Minnen}, \bibinfo{person}{Eirikur Agustsson}, {and} \bibinfo{person}{Michael Tschannen}.} \bibinfo{year}{2023}\natexlab{}.
\newblock \showarticletitle{Finite scalar quantization: Vq-vae made simple}.
\newblock \bibinfo{journal}{\emph{arXiv preprint arXiv:2309.15505}} (\bibinfo{year}{2023}).
\newblock


\bibitem[Pi et~al\mbox{.}(2019)]%
        {pi2019practice}
\bibfield{author}{\bibinfo{person}{Qi Pi}, \bibinfo{person}{Weijie Bian}, \bibinfo{person}{Guorui Zhou}, \bibinfo{person}{Xiaoqiang Zhu}, {and} \bibinfo{person}{Kun Gai}.} \bibinfo{year}{2019}\natexlab{}.
\newblock \showarticletitle{Practice on long sequential user behavior modeling for click-through rate prediction}. In \bibinfo{booktitle}{\emph{Proceedings of the 25th ACM SIGKDD international conference on knowledge discovery \& data mining}}. \bibinfo{pages}{2671--2679}.
\newblock


\bibitem[Pi et~al\mbox{.}(2020)]%
        {pi2020search}
\bibfield{author}{\bibinfo{person}{Qi Pi}, \bibinfo{person}{Guorui Zhou}, \bibinfo{person}{Yujing Zhang}, \bibinfo{person}{Zhe Wang}, \bibinfo{person}{Lejian Ren}, \bibinfo{person}{Ying Fan}, \bibinfo{person}{Xiaoqiang Zhu}, {and} \bibinfo{person}{Kun Gai}.} \bibinfo{year}{2020}\natexlab{}.
\newblock \showarticletitle{Search-based user interest modeling with lifelong sequential behavior data for click-through rate prediction}. In \bibinfo{booktitle}{\emph{Proceedings of the 29th ACM International Conference on Information \& Knowledge Management}}. \bibinfo{pages}{2685--2692}.
\newblock


\bibitem[Radford et~al\mbox{.}(2021)]%
        {radford2021learning}
\bibfield{author}{\bibinfo{person}{Alec Radford}, \bibinfo{person}{Jong~Wook Kim}, \bibinfo{person}{Chris Hallacy}, \bibinfo{person}{Aditya Ramesh}, \bibinfo{person}{Gabriel Goh}, \bibinfo{person}{Sandhini Agarwal}, \bibinfo{person}{Girish Sastry}, \bibinfo{person}{Amanda Askell}, \bibinfo{person}{Pamela Mishkin}, \bibinfo{person}{Jack Clark}, {et~al\mbox{.}}} \bibinfo{year}{2021}\natexlab{}.
\newblock \showarticletitle{Learning transferable visual models from natural language supervision}. In \bibinfo{booktitle}{\emph{International conference on machine learning}}. PmLR, \bibinfo{pages}{8748--8763}.
\newblock


\bibitem[Rajput et~al\mbox{.}(2023)]%
        {rajput2023recommender}
\bibfield{author}{\bibinfo{person}{Shashank Rajput}, \bibinfo{person}{Nikhil Mehta}, \bibinfo{person}{Anima Singh}, \bibinfo{person}{Raghunandan Hulikal~Keshavan}, \bibinfo{person}{Trung Vu}, \bibinfo{person}{Lukasz Heldt}, \bibinfo{person}{Lichan Hong}, \bibinfo{person}{Yi Tay}, \bibinfo{person}{Vinh Tran}, \bibinfo{person}{Jonah Samost}, {et~al\mbox{.}}} \bibinfo{year}{2023}\natexlab{}.
\newblock \showarticletitle{Recommender systems with generative retrieval}.
\newblock \bibinfo{journal}{\emph{Advances in Neural Information Processing Systems}}  \bibinfo{volume}{36} (\bibinfo{year}{2023}), \bibinfo{pages}{10299--10315}.
\newblock


\bibitem[Si et~al\mbox{.}(2024)]%
        {si2024twin}
\bibfield{author}{\bibinfo{person}{Zihua Si}, \bibinfo{person}{Lin Guan}, \bibinfo{person}{ZhongXiang Sun}, \bibinfo{person}{Xiaoxue Zang}, \bibinfo{person}{Jing Lu}, \bibinfo{person}{Yiqun Hui}, \bibinfo{person}{Xingchao Cao}, \bibinfo{person}{Zeyu Yang}, \bibinfo{person}{Yichen Zheng}, \bibinfo{person}{Dewei Leng}, {et~al\mbox{.}}} \bibinfo{year}{2024}\natexlab{}.
\newblock \showarticletitle{Twin v2: Scaling ultra-long user behavior sequence modeling for enhanced ctr prediction at kuaishou}. In \bibinfo{booktitle}{\emph{Proceedings of the 33rd ACM International Conference on Information and Knowledge Management}}. \bibinfo{pages}{4890--4897}.
\newblock


\bibitem[Team et~al\mbox{.}(2025)]%
        {team2025kwai}
\bibfield{author}{\bibinfo{person}{Kwai~Keye Team}, \bibinfo{person}{Biao Yang}, \bibinfo{person}{Bin Wen}, \bibinfo{person}{Changyi Liu}, \bibinfo{person}{Chenglong Chu}, \bibinfo{person}{Chengru Song}, \bibinfo{person}{Chongling Rao}, \bibinfo{person}{Chuan Yi}, \bibinfo{person}{Da Li}, \bibinfo{person}{Dunju Zang}, {et~al\mbox{.}}} \bibinfo{year}{2025}\natexlab{}.
\newblock \showarticletitle{Kwai Keye-VL Technical Report}.
\newblock \bibinfo{journal}{\emph{arXiv preprint arXiv:2507.01949}} (\bibinfo{year}{2025}).
\newblock


\bibitem[Wang et~al\mbox{.}(2024)]%
        {home}
\bibfield{author}{\bibinfo{person}{Xu Wang}, \bibinfo{person}{Jiangxia Cao}, \bibinfo{person}{Zhiyi Fu}, \bibinfo{person}{Kun Gai}, {and} \bibinfo{person}{Guorui Zhou}.} \bibinfo{year}{2024}\natexlab{}.
\newblock \showarticletitle{HoME: Hierarchy of Multi-Gate Experts for Multi-Task Learning at Kuaishou}.
\newblock  (\bibinfo{year}{2024}).
\newblock


\bibitem[Wu et~al\mbox{.}(2025)]%
        {wu2025muse}
\bibfield{author}{\bibinfo{person}{Bin Wu}, \bibinfo{person}{Feifan Yang}, \bibinfo{person}{Zhangming Chan}, \bibinfo{person}{Yu-Ran Gu}, \bibinfo{person}{Jiawei Feng}, \bibinfo{person}{Chao Yi}, \bibinfo{person}{Xiang-Rong Sheng}, \bibinfo{person}{Han Zhu}, \bibinfo{person}{Jian Xu}, \bibinfo{person}{Mang Ye}, {et~al\mbox{.}}} \bibinfo{year}{2025}\natexlab{}.
\newblock \showarticletitle{MUSE: A Simple Yet Effective Multimodal Search-Based Framework for Lifelong User Interest Modeling}.
\newblock \bibinfo{journal}{\emph{arXiv preprint arXiv:2512.07216}} (\bibinfo{year}{2025}).
\newblock


\bibitem[Yang et~al\mbox{.}(2025a)]%
        {yang2025qwen3}
\bibfield{author}{\bibinfo{person}{An Yang}, \bibinfo{person}{Anfeng Li}, \bibinfo{person}{Baosong Yang}, \bibinfo{person}{Beichen Zhang}, \bibinfo{person}{Binyuan Hui}, \bibinfo{person}{Bo Zheng}, \bibinfo{person}{Bowen Yu}, \bibinfo{person}{Chang Gao}, \bibinfo{person}{Chengen Huang}, \bibinfo{person}{Chenxu Lv}, {et~al\mbox{.}}} \bibinfo{year}{2025}\natexlab{a}.
\newblock \showarticletitle{Qwen3 technical report}.
\newblock \bibinfo{journal}{\emph{arXiv preprint arXiv:2505.09388}} (\bibinfo{year}{2025}).
\newblock


\bibitem[Yang et~al\mbox{.}(2025b)]%
        {yang2025kwai}
\bibfield{author}{\bibinfo{person}{Biao Yang}, \bibinfo{person}{Bin Wen}, \bibinfo{person}{Boyang Ding}, \bibinfo{person}{Changyi Liu}, \bibinfo{person}{Chenglong Chu}, \bibinfo{person}{Chengru Song}, \bibinfo{person}{Chongling Rao}, \bibinfo{person}{Chuan Yi}, \bibinfo{person}{Da Li}, \bibinfo{person}{Dunju Zang}, {et~al\mbox{.}}} \bibinfo{year}{2025}\natexlab{b}.
\newblock \showarticletitle{Kwai keye-vl 1.5 technical report}.
\newblock \bibinfo{journal}{\emph{arXiv preprint arXiv:2509.01563}} (\bibinfo{year}{2025}).
\newblock


\bibitem[Yang et~al\mbox{.}(2020)]%
        {yang2020large}
\bibfield{author}{\bibinfo{person}{Xiaoyong Yang}, \bibinfo{person}{Yadong Zhu}, \bibinfo{person}{Yi Zhang}, \bibinfo{person}{Xiaobo Wang}, {and} \bibinfo{person}{Quan Yuan}.} \bibinfo{year}{2020}\natexlab{}.
\newblock \showarticletitle{Large scale product graph construction for recommendation in e-commerce}.
\newblock \bibinfo{journal}{\emph{arXiv preprint arXiv:2010.05525}} (\bibinfo{year}{2020}).
\newblock


\bibitem[Zhou et~al\mbox{.}(2025)]%
        {zhou2025onerec}
\bibfield{author}{\bibinfo{person}{Guorui Zhou}, \bibinfo{person}{Jiaxin Deng}, \bibinfo{person}{Jinghao Zhang}, \bibinfo{person}{Kuo Cai}, \bibinfo{person}{Lejian Ren}, \bibinfo{person}{Qiang Luo}, \bibinfo{person}{Qianqian Wang}, \bibinfo{person}{Qigen Hu}, \bibinfo{person}{Rui Huang}, \bibinfo{person}{Shiyao Wang}, {et~al\mbox{.}}} \bibinfo{year}{2025}\natexlab{}.
\newblock \showarticletitle{OneRec Technical Report}.
\newblock \bibinfo{journal}{\emph{arXiv preprint arXiv:2506.13695}} (\bibinfo{year}{2025}).
\newblock


\bibitem[Zhou et~al\mbox{.}(2018)]%
        {zhou2018deep}
\bibfield{author}{\bibinfo{person}{Guorui Zhou}, \bibinfo{person}{Xiaoqiang Zhu}, \bibinfo{person}{Chenru Song}, \bibinfo{person}{Ying Fan}, \bibinfo{person}{Han Zhu}, \bibinfo{person}{Xiao Ma}, \bibinfo{person}{Yanghui Yan}, \bibinfo{person}{Junqi Jin}, \bibinfo{person}{Han Li}, {and} \bibinfo{person}{Kun Gai}.} \bibinfo{year}{2018}\natexlab{}.
\newblock \showarticletitle{Deep interest network for click-through rate prediction}. In \bibinfo{booktitle}{\emph{Proceedings of the 24th ACM SIGKDD international conference on knowledge discovery \& data mining}}. \bibinfo{pages}{1059--1068}.
\newblock


\end{thebibliography}
\end{document}